\documentstyle[aps,multicol,psfig]{revtex}

\begin{document}
\draft
\title{Ratchet effect in  surface electromigration: \\
       smoothing surfaces by an ac field}
\author{Imre Der\'enyi$^1$, Choongseop Lee$^2$ and Albert-L\'aszl\'o Barab\'asi$^{2,*}$}
\address{$^1$Department of Atomic Physics, E\"otv\"os University,
         Budapest, Puskin u 5-7, 1088 Hungary \\
         $^2$Department of Physics, University of Notre Dame,
         Notre Dame, IN 46556}
\date{Received 14 August 1997}

\maketitle

\begin{abstract}
We demonstrate that for surfaces that have a nonzero Schwoebel 
barrier the application of an ac field parallel to the surface 
induces a net electromigration current, that points in the 
descending step direction. The magnitude of the current is 
calculated analytically and compared with Monte Carlo simulations. Since a  
downhill current smoothes the surface, our results imply that 
the application of ac fields can aid the smoothing 
process during annealing and can slow or eliminate 
the Schwoebel barrier induced mound formation during growth.
\end{abstract}


\begin {multicols}{2}
\narrowtext

\begin{figure}[htb]
\begin{center}
\hskip 2.0 cm
\mbox{\psfig{figure=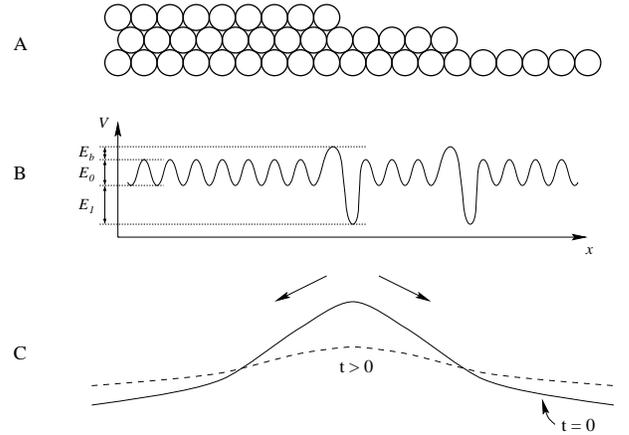,width=8 cm}}
\end{center}
\caption{
Schematic illustration of the cross section of a
vicinal
surface containing two monatomic steps ({\bf A}) and the asymmetric
lattice potential ({\bf B}) experienced by an atom diffusing on this surface. Note that if the Schwoebel barrier $E_b=0$, the lattice potential near a step is spatially symmetric, while this symmetry is broken for $E_b \neq 0$. A downhill (uphill) current points to the right (left) on this figure. As ({\bf C}) illustrates, if there is a mound on the surface at $t=0$, a downhill current indicated by the arrows will tend to decrease its height, i.e. it smoothes the film by the reorganization of the material on the surface.}
\label{fig1}
\end{figure}

Growing epitaxial films with smooth surfaces is one of the ongoing challenges of the thin film community. However, this goal is hampered by a series of basic physical effects  that lead to the development of unavoidable surface roughness during growth. In particular, there is abundant experimental and theoretical evidence that during
deposition the diffusion bias generated by the Schwoebel barrier
(see Fig.\ \ref{fig1}) results in a net {\it uphill current}, which in turn leads to the
formation of mounds and to a fast and unwanted increase in the
interface roughness \cite{johnson}. As Fig.\ \ref{fig1} demonstrates, the Schwoebel barrier introduces spatial asymmetry in the otherwise symmetric lattice potential.  Interestingly, in the recent years has been recognized that in such periodic and spatially asymmetric systems (ratchets)
  non-equilibrium fluctuations can {\it
rectify}
Brownian motion  and induce a  nonzero net current \cite{ratchet}.  This
{\it fluctuation driven transport} is believed to play an essential role in the
operation of motor proteins or molecular motors, and might result in new separation techniques \cite{appl}. 
In this paper we propose the first  nano-scale application of this
transport
mechanism  based on the atomic {electromigration}
on
vicinal surfaces induced by alternating electric
fields. We demonstrate that the Schwoebel barrier induced asymmetry in the lattice potential can be used to generate a {\it downhill current}, aiding the smoothing of surfaces during growth or annealing.

Atom diffusion on crystal surfaces  is a thermally activated process:
 atoms can hop from their position to a neighboring
one by overcoming a potential barrier $\Delta E$.  The hopping rate is given
by the Arrhenius law
 $k = \nu_0 \exp(-{ \Delta E / k_{\rm B} T }) $,
where $T$ is the temperature and $\nu_0$ is the vibration frequency of the surface atoms.
 Fig.\ \ref{fig1}
illustrates the lattice potential
of a vicinal surface that consists of long flat terraces
separated
by
monatomic steps. The barrier height for  diffusion on a
flat surface is
denoted
by $E_0$. Near a step atoms form additional lateral bonds of energy $E_1$ with the step atoms, leading to a  deeper  potential
valley.  Finally, jumping  over a
step requires passing an  additional potential barrier, the Schwoebel
barrier,
$E_{\rm b}$ \cite{schwoebel}.

For most  metals and semiconductors the 
otherwise random surface diffusion  of
the atoms can be  biased by an external electric field
applied
parallel to the surface, a phenomenon known as surface
electromigration \cite{yasunaga}. The effective force, {\boldmath $F$}, acting
on the surface  atoms is proportional to the field {\boldmath
$E$},
{\boldmath $F$} $= Z e${\boldmath $E$},
where $e$ ($>0$) is the elementary charge and $Z$ is the
effective
charge number which consists of two terms, $Z = Z_{\rm d} +
Z_{\rm
w}$.

The ``direct'' term $Z_{\rm d}$ ($>0$) is associated with the
electrostatic interaction between the atom and the electric
field, while the ``wind''
term
$Z_{\rm w}$ ($<0$) is generated by the scattering of the current
carrying electrons on the surface atoms. The competition between  these
two terms can result in either positive or negative effective
charge \cite{kandel}.

The constant electric field induces a  surface  current parallel to  the field.
However, it is expected that  a
{\it periodic field} with zero mean  does {\it not} create a net current
over a full period, since it simply amounts to biasing in an
equal manner the motion of the atoms in two opposing  directions. 
In contrast, next we
demonstrate that for systems that have a nonzero Schwoebel barrier the asymmetry induced by the barrier rectifies the diffusion process, 
generating a net current along the surface, even if the external
field has a zero mean value. Most important, the induced current is {\it always  downhill}, i.e. it points towards the
descending step direction,  independent of the step orientation
or the effective charge.  Since the downhill  current acts to smooth the surface, it has the potential to accelerate the smoothing process during annealing and to slow or eliminate the
Schwoebel barrier induced mound formation during growth. Consequently,  this nano-scale ratchet effect can have important
technological applications for thin film growth.

\begin{figure}[htb]
\begin{center}
\hskip -3.0 cm
\mbox{\psfig{figure=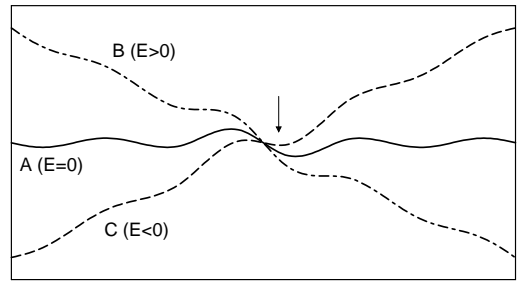,width=3.5 cm,angle=-90}}
\end{center}
\vskip -1.5 cm
\caption{The effect of an uniform electric field on the potential landscape experienced by an atom diffusing on the surface. The line ({\bf A}) corresponds to the lattice potential in the vicinity of a step for $E=0$ (see also Fig.\ \ref{fig1}B). Assuming $Z > 0$, a large enough field to the right ($E > 0$) 'tilts' the potential landscape, eliminating  the  minima of the potential,  
 biasing the adatom motion to the right (see ({\bf B})). On the other hand, as ({\bf C}) illustrates, an inverse field ($E < 0$) of the same magnitude does not eliminate the minima in the vicinity  of the step (marked by an arrow), locally trapping the particle. Consequently, 
the current to the left in ({\bf C}) is smaller than the current to the right in ({\bf B}), and a periodic external field will result in a net current to the right, i.e. a downhill current. For smaller fields there are minima 
in both cases, but the minimum corresponding to ({\bf C}) is deeper than 
that of ({\bf B}), again resulting in a more efficient trapping of the atoms
moving uphill, and to  a net downhill current. Note that the direction of the net current is determined only by the step orientation, and not by the sign of $Z$.} 
\label{fig2}\end{figure}

Consider  a vicinal semiconductor surface with terraces of
equal
width $w$ in an alternating external electric field (with
zero
mean) perpendicular to the steps and parallel to the
surface. The lattice potential of this system is periodic
and
{\it spatially asymmetric}, and the alternating field acts as a zero-mean non-equilibrium fluctuating force. In the presence of such field (or ac current) a
directed
net flow of the diffusing atoms on the surface is expected from
the
theory of fluctuation driven transport \cite{ratchet}. A simple explanation of the effect responsible for this current is given in Fig.\ \ref{fig2}. 
To evaluate the net  current we 
first
consider the motion of a {\it single} atom 
on a {\it fixed} vicinal surface with terraces of width
$w$.
Since the diffusion parallel to the steps is not affected by the
electric field we can neglect  the  transverse
direction and
consider
the motion in one dimension, perpendicular to the steps. The one-dimensional
lattice
potential for the diffusing atom  is similar to the potential
of
Fig.\ \ref{fig1}, but it is periodic with period $w$, each
period
 containing  $\ell$ valleys and barriers, where $\ell=w/a$
is
the dimensionless size of the terraces measured in the units of
the
lattice constant $a$.  Associated with the steps, every ($\ell-1$)th $E_0$ barrier is followed by a higher  Schwoebel barrier $E_b$ and  a deeper valley $E_1$ representing binding to the step. The system can be reduced to one period of the potential
with
periodic boundary conditions, in which the motion of the atom is
described by the master equation

\begin{equation}
 {\partial P_i(t) \over \partial t} =
  k^+_{i-1}(t) P_{i-1}(t) + k^-_{i+1}(t) P_{i+1}(t) - (k^+_i(t) + k^-_i(t))
P_i(t),
\label{eq_master}
\end{equation}
where $1\leq i \leq \ell$ with periodic boundary conditions, $P_i(t)$
denotes the probability that the atom is situated in the $i$th
valley
at time $t$.  $k^-_i$ and $k^+_i$ denote the hopping rates
from
the
$i$th valley to the left and to the right, respectively, where 
 $k^-_i(t) =
  \nu_0 \, {\rm e}^{-\left[\Delta E^-_i + \delta E(t)\right] / k_{\rm B}T}$, and
 $k^+_i(t) =
  \nu_0 \, {\rm e}^{-\left[\Delta E^+_i - \delta E(t)\right] / k_{\rm B}T}$,
$\Delta E^-_1 = E_0$,\ \ $\Delta E^+_1 = E_0+E_{\rm b}$,\ \
$\Delta E^-_2 = E_0+E_1+E_{\rm b}$,\ \ $\Delta E^+_2 = E_0+E_1$, and
$\Delta E^-_j = \Delta E^+_j = E_0$ for $3 \leq j \leq \ell$
are the heights of the barriers of the lattice potential. Finally,
$\delta E(t) = F(t) a/2$ is the change in the barrier heights
associated with the force $F(t)=ZeE(t)$ induced by the electric field
$E(t)$.

If the period of the electric field is much larger than the relaxation time $\tau \approx \ell^2
\exp(E_0/k_{\rm B}T)/\nu_0$ of the probability distribution $P_i(t)$, $P_i(t)$ is ``slaved'' by the electric field and
the solution of the master equation reduces to the
stationary solution with  constant electric field $E$. If the period of the field is much smaller than $\tau$, $P_i(t)$ has no time to adjust to the field, resulting in a very small net current. Thus $f_u = 1/\tau$ is the upper limit
of the frequency, since exceeding it the current goes to
zero. Therefore, in the following we assume that the frequency $f$ of
the alternating field is below the upper limit $f_u$, so that the
stationary solution of the master equation applies.  In this case  Eq.\ (\ref{eq_master}) reduces to a system of ($\ell-1$) independent linear
equations with the normalisation  condition $\sum_{i=1}^\ell P_i = 1$,  giving the
$\ell$th equation. The solution provides 
the stationary current as 
\begin{eqnarray}
 && J_{\rm st}(F) = k^+_1(F) P_1(F) - k^-_2(F) P_2(F) =
\nonumber\\
 && \qquad\quad
  = 2\,\sinh(\ell\delta E/k_{\rm B}T)\,\nu_0\,{\rm e}^{-E_0/k_{\rm B}T}
 \times \nonumber \\ && \times
 1 / \biggl\{
     \left({\rm e}^{E_b/k_{\rm B}T}-1\right)
     \left({\rm e}^{E_1/k_{\rm B}T}-1\right)
     {\rm e}^{(1-\ell)\delta E/k_{\rm B}T} +
\nonumber\\
 && +
     \left[\ell+\left({\rm e}^{E_b/k_{\rm B}T}-1\right)
     +\left({\rm e}^{E_1/k_{\rm B}T}-1\right)\right]
     \frac{\sinh(\ell\delta E/k_{\rm B}T)}{\sinh(\delta E/k_{\rm B}T)}
  \biggr\} \; .
\nonumber\\
\label{eq_Jst}
\end{eqnarray}

For simplicity we restrict the calculation to the case when the field
is as a symmetric square wave, i.e. the force alternates
between $+F$ and $-F$ at constant intervals.  In this case, when  $ f \ll f_u$, the net current $J_0$ can be calculated from (\ref{eq_Jst}) as $J_0=[J_{\rm st}(+F)+J_{\rm st}(-F)]/2$.

If the interaction between the diffusing atoms is neglected, the
average number of atoms that are able to move (i.e., that are not
part of a terrace) is $1+N_a$ in each period. The first term indicates
the edge of the terrace and the second term,
$N_a=(\ell-1)/[1+\exp(E_1/k_{\rm B}T)]$,
is the average number of diffusing atoms (adatoms) on the surface.  Thus the net particle current is
\begin{equation}
 J = J_{0}(F) (1+N_a) =
     J_{0}(F) \biggl( 1 + {\ell-1 \over 1+{\rm e}^{E_1/k_{\rm B}T} } \biggr) \; .
\label{eq_J}
\end{equation}

\begin{figure}[htb]
\begin{center}
\vskip -1.3 cm
\hskip -3.0 cm
\mbox{\psfig{figure=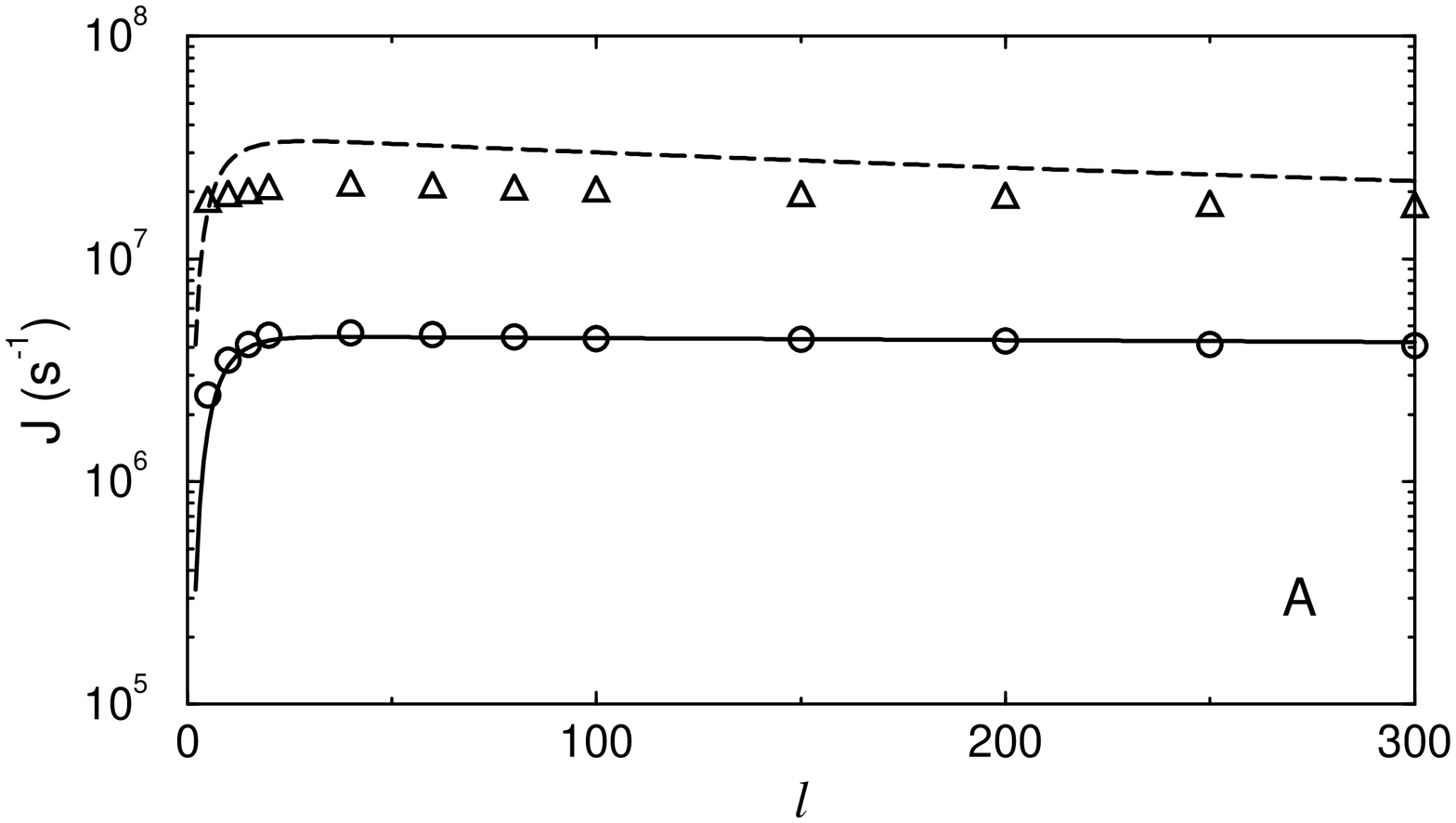,width=3.5 cm,angle=-90}}
\vskip -2.0 cm
\hskip -3.0 cm
\mbox{\psfig{figure=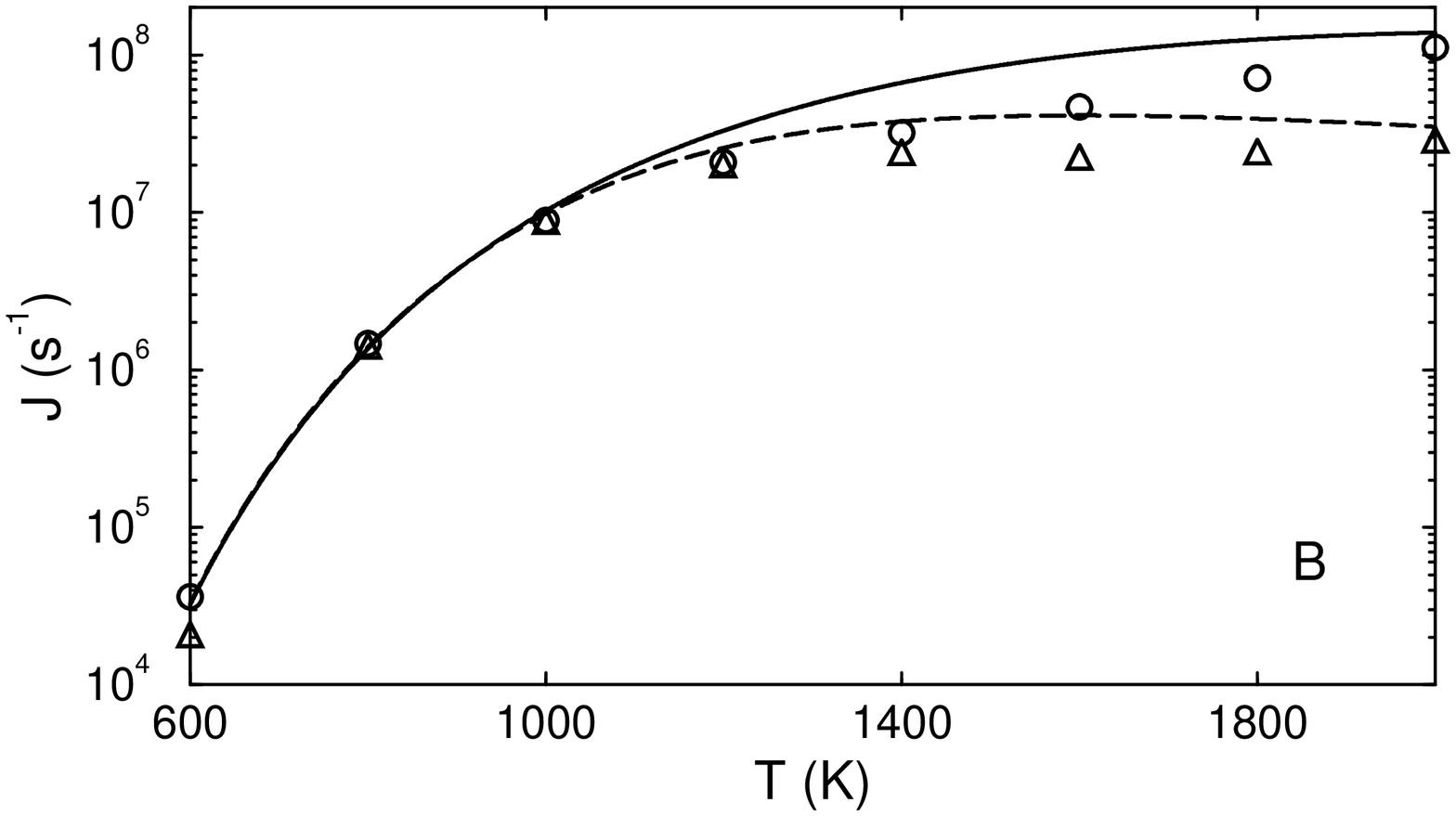,width=3.5 cm,angle=-90}}
\end{center}
\vskip -1.0 cm
\caption{({\bf A}) The ac field induced downhill current as a function if 
the average step size $\ell$. The lines correspond to the analytical results (\ref{eq_Jst}) and (\ref{eq_J}) for T=900 K (continuous line) and T=1200 K 
(dashed line), while the circles (900 K) and triangles (1200 K) are the 
currents predicted by the MC simulations. ({\bf B}) The temperature 
dependence of the net current for average step sizes $\ell=20$  and  
$\ell=200$. The  continuous ($\ell=20$) and dashed  ($\ell=200$) 
lines correspond to the analytical predictions, while the circles 
($\ell=20$) and triangles ($\ell=200$) are the result of the MC
simulations. In the simulations we used the parameters $E_0=0.3$ eV, 
$E_1=0.6$ eV, $E_b=0.15$ eV, $Z=0.5$,  $E=10^8$ V/m, and $\nu_o=10^{13}$ s$^{-1}$. The system size is $L=2000$ and the results were averaged over 20 independent runs.} 
\label{fig3}\end{figure}

Eq.\ (\ref{eq_J}) provides the downhill current generated by the interplay between the ac field and the Schwoebel barrier. The terrace size dependence of this current for two different temperatures is shown in Fig.\ \ref{fig3}A.

The previous calculation, while  correctly describes the nature
and the qualitative features of the net current, neglects the   atom-atom interaction and the step fluctuations. Since the
source of atoms are the steps (adatoms detach from step edges), the step length is not fixed, but it
fluctuates. To incorporate these
effects and to check the validity of the analytical prediction, we  performed Monte Carlo (MC) simulations
with activated diffusion along the surface.  In the simulations we
 start with a series of steps of length $\ell$. Every surface
atom that has less than
two  neighbors in the same layer is allowed to diffuse with a
probability
$P \sim \exp(- \Delta E/k_BT)$. In the presence
of an  electric field $\delta E$ the activation energy is reduced
or
increased with  $\delta E$, depending on whether the direction of
the hop is 
along or against the field.  In the simulations we use
a  periodic square wave with frequency $ f \ll f_u$, and determine the net current defined as the number of excess hops in the downhill direction in
the unit time,  normalized to the system size.

As Fig.\ \ref{fig3}A indicates,  the net current obtained in the Monte Carlo simulation qualitatively agrees with the prediction (\ref{eq_J}) of the
master equation. Furthermore, there is excellent quantitative agreement at low temperature (900 K) while (\ref{eq_J}) overestimates the current at higher $T$. Note that the fit does not require any fitting parameter.

To pinpoint the temperature range for which the analytical solution provides a good approximation, in Fig.\ \ref{fig3}B we plot the temperature dependence of the net current for $\ell=20$ and $\ell=200$. We find excellent agreement between the theory and the MC simulations for $T < T^*$ ($\simeq$ 1000 K), while for $T > T^*$ the current predicted by (\ref{eq_J}) systematically exceeds the numerical result. Indeed, increasing the temperature  also increases the adatom density on the surface, and consequently the role of the neglected  atom-atom interactions  also increases \cite{derenyi}. The value of $T^*$ depends only on the energy barriers $E_0, E_1$ and $E_b$. In some materials  the bonding energies might be  higher than the quoted values, which further increases $T^*$ and the range of applicability of the analytical solution. Note that the agreement between the analytical predictions and the MC simulations is better for $\ell=200$ than $\ell=20$, underlying the larger impact of step fluctuation on shorter steps \cite{X}.

Since in the typical electromigration experiments \cite{yasunaga} $\delta E$ is much smaller than $k_{B}T$, we can expand the current $J_{0}(F)$ into Taylor series in terms of $(\delta E/k_{B}T)$, obtaining that the net current is a second order effect, proportional to $(Z\delta E)^{2}$. The MC simulations confirm this prediction, providing a quantitative expression for tuning the current with $E$ \cite{comment1}.    

In conclusion, we have  demonstrated  both analytically and numerically that  the Schwoebel barrier, in the presence of a periodic external electric field, leads to a downhill current. Since most metal and semiconductor surfaces do have  a nonzero Schwoebel barrier and display electromigration, we expect that the appearance of such a net current is relevant for a large class of technologically important materials. Thus the application of an ac current during either growth or annealing can lead to a nontrivial  smoothing effect, and aid the growth of smooth surfaces. This consequence of the ratchet effect can thus have important practical applications in the growth and processing of high quality thin films. 

\vskip 1.5 cm
\footnotesize
$^*$Corresponding authors

Electronic address: alb@nd.edu

\end{multicols}


\begin{references}
\bibitem{X} When a small step emits (absorbs) an atom, its relative change in size will be much larger than that of a larger step.  Thus one expects a better agreement  between the theory and the simulation for large $\ell$.  On the other hand, fluctuation in the step sizes for small $\ell$ are smaller due to step-step interaction, balancing this effect.
\bibitem{comment1} In order to amplify the effect in the timespan allowed by the MC simulation, in Fig. \ \ref{fig3} we used $ZE=5\times10^7$ V/m, larger than the typical values used in current experiments \cite{yasunaga}, that are of order $10^5$ V/m. However, using the expression for the field dependence, we can calculate the current for arbitrary electric fields.
\end{references}
\end{document}